\documentstyle[preprint,aps]{revtex}
\begin{document}
\tighten
\title{A method of calculating  the Jost function for
 analytic potentials}
   \author{
    S. A. Rakityansky\cite{Dubna},
 S. A.  Sofianos, and K. Amos\cite{Melb}}
\address{ Physics Department, University of South Africa,  P.O.Box
     392, Pretoria 0001, South Africa}
\date{\today}
\maketitle
\begin{abstract}
A combination of the variable-constant and complex coordinate rotation
methods is used to solve the two-body Schr\"odinger equation. The latter is
replaced by a system of linear first-order differential equations, which
enables one to perform direct calculation of the Jost function for all
complex momenta of physical interest, including the spectral points
corresponding to bound and resonance states. Explicit forms of the equations,
appropriate for central, noncentral, and Coulomb-tailed potentials are given.
Within the proposed method, the scattering, bound, virtual, and resonance
state problems can be treated in a unified way. The effectiveness of the
method is demonstrated by a numerical example.\\
{PACS numbers: 03.65.Nk, 21.45.+v, 02.70.+d}
\end{abstract}


\section{Introduction}
The nonrelativistic quantum mechanical two-body problem can be described
conveniently and nicely
 in terms of the Jost functions and Jost solutions of the
Schr\"odinger equation. When defined for all complex values of the
momentum, these functions contain complete information about the underlying
physical system. In contrast to the conventional treatment of the two-body
problem, based on the scattering amplitude and physical wave-function,
in the Jost function approach
the bound, virtual, scattering, and resonance states are
treated on an equal footing and simultaneously. Also the Jost function is a
more fundamental quantity than the $S$ function  since it is
free of ambiguities caused by  redundant zeros, while
the $S$ function  may  have redundant poles\cite{rpol}.

Despite the fact that the Jost functions are more
appropriate, in principle, for studies of
spectral properties of two-body Hamiltonians, they  have not been
used in the past in  practical calculations.
 Instead they were considered as
purely mathematical entities  of  formal scattering theory,
which can be  expressed
either via an integral containing the regular solution or
via a Wronskian of the regular and Jost solutions\cite{newton}. Thus,
to find  the Jost function based upon such information, the
 full solution of the problem itself is required.

In Ref.~\cite{rakpup}, linear first-order differential equations for
functions closely related to the Jost solutions were proposed.
 These equations are based on the variable-phase approach\cite{calog} and
their solution at any fixed value of the radial variable $r$, provides the
Jost function and its complex conjugate counterpart, which correspond to the
potential truncated at the point $r$. The proposed  method, however,  is
suitable  only for bound and scattering states calculations, that is for
calculations in the upper half of the complex momentum plane.

Herein we extend  the method of  Ref.~\cite{rakpup} onto the unphysical sheet
so that the resonance state region is included.  We do so by a combination
of the variable-constant method\cite{vcm} and the
 `complex coordinate rotation' method\cite{ccr}.
The latter method, often used in studies of atomic and molecular physics,
is also known as the 'complex-scaling' or the `complex-dilation' method.
It is based on the fact that the complex
coordinate transformation, $\vec r\to {\vec r} \exp(i \theta)$ with a
sufficiently large $\theta > 0$, makes the resonance state wave function
quadratically integrable leaving the energies and widths of the bound and
resonance states  unaffected since
the Jost function, and  the positions of its zeros in particular, do not
depend on that vector $\vec r$. This is  true if the potential
is an analytic function of the
coordinate $r$ and vanishes rapidly enough when that coordinate
goes to infinity along the turned ray\cite{newton}.
In atomic and molecular physics, this property has been used to locate
resonances by the same variational methods employed  for the bound
states\cite{kuk}.

In this paper we make a complex rotation of the radial variable in the
differential equations proposed in Ref.~\cite{rakpup}. The new
equations that result, facilitate evaluation of the Jost function
in the entire complex momentum plane and therefore one can locate  the
resonances as the zeros of the Jost function.

The paper is organized as follows. In Sec.~II we derive the rotated
equations in the simplest case of a central short--range potential. In
Sec.~III the equations are generalised for the case of a noncentral
potential, while  Sec.~IV is devoted to consideration of
 potentials having a Coulomb tail. In Sec.~V
the method is tested by comparing results with the properties of an exactly
solvable model. Finally, possible further extensions of the method, its
advantages and drawbacks are discussed in Sec.~VI.


\section{Central short--range potential}

Consider the  problem of two quantal particles interacting
via a central potential $V(r)$ with the asymptotic properties
\begin{equation}
\label{1}
      \lim_{r\to 0} r^2 V(r) = 0\ ,
\end{equation}
and
\begin{equation}
\label{2}
	\lim_{r\to \infty} r^{1+\varepsilon} V(r) = 0\ ,
		   \qquad \varepsilon>0 \ .
\end{equation}
Any physical solution of the radial Schr\"odinger equation ($\hbar=2m=1$)
\begin{equation}
\label{3}
       D_{\ell}(k,r)\ \Phi_{\ell}(k,r) = V(r)\ \Phi_{\ell}(k,r)\ ,
\end{equation}
where
\begin{equation}
     D_{\ell}(k,r)\ \equiv \  \partial_r^2 + k^2 - \ell(\ell+1)/r^2\ ,
\end{equation}
is proportional to the so-called regular solution which is defined by the
boundary condition at the origin,
\begin{equation}
\label{5}
     \lim_{r\to 0}\left [\Phi_\ell(k,r)/\jmath_\ell(kr)\right ] = 1\ .
\end{equation}
Here $\jmath_\ell(z)$ is the Riccati-Bessel function\cite{abram}.
At   large distances the regular solution is a linear combination
\cite{taylor} of Riccati-Hankel functions $h_\ell^{(\pm)}(kr)$ \cite{abram},
 i.e.
\begin{equation}
\label{6}
   \Phi_\ell(k,r)\ \mathop{\longrightarrow}\limits_{r\to\infty}\ \frac{1}{2}
      \left[h_\ell^{(+)}(kr)\ f_\ell^*(k^*) +
    h_\ell^{(-)}(kr)\ f_\ell(k)\right]
  \ ,
\end{equation}
wherein the $r-$independent quantity $f_\ell(k)$ is the Jost function whose
zeros in the complex $k$--plane correspond to the bound, ($R\!e\,
k=0,\,I\!m\,k > 0$ ), virtual, ($R\!e\, k=0,\,I\!m\,k<0$ ), and
resonance, ($R\!e\,k>0,\,I\!m\,k<0$ ), states of the system.

Suppose $V(r)$ is an analytic potential\cite{newton}, i.e. it can be
defined as an analytic function of the complex variable
\begin{equation}
\label{7}
      r=x \exp(i\theta),\qquad x\ge0,
\qquad \theta\in[0,\theta_{max}],\quad\theta_{max}<\pi/2\ ,
\end{equation}
and  fulfills the conditions of Eqs.~(\ref{1}) and (\ref{2}).
We consider the complex dilation of
the regular solution using this rotated coordinate. The corresponding
function $\Phi_\ell(k,x e^{i\theta})$ is square integrable over the
variable $x$, even for a resonance state, provided the angle $\theta$ is
greater than  $\varphi$ which defines the position of the resonance on the
lower half of the complex $k$-plane by
\begin{equation}
\label{7*}
	  k_0=|k_0|\exp(-i\varphi) .
\end{equation}
 This is obvious because
the product `$k_0r$' in the asymptotic behaviour $\sim \exp(ik_0r)$ of the
solution has a positive imaginary part.

To obtain the equation for the dilated regular solution, we substitute
Eq.~(\ref{7}) into Eqs.~(\ref{3}) and (\ref{5}) giving
\begin{equation}
\label{8}
     D_\ell(ke^{i\theta},x)\ \Phi_\ell(k,xe^{i\theta})  =
	  e^{2i\theta}V(xe^{i\theta})\ \Phi_\ell(k,xe^{i\theta})\ ,
\end{equation}
 and
\begin{equation}
\label{9}
   \lim_{x\to 0}   \left [
	\Phi_\ell(k,xe^{i\theta})\left/j_\ell(kxe^{i\theta})
	       \right.   \right ]=1 \ .
\end{equation}
Furthermore, and since  the potential is assumed to be analytic,
substitution of Eq.~(\ref{7}) into Eq.~(\ref{6}) gives  the asymptotic form
\begin{equation}
\label{10}
    \Phi_\ell(k,xe^{i\theta})\  \mathop{\longrightarrow}
		\limits_{x\to\infty}\
	       \frac{1}{2}\left [ h_\ell^{(+)}(kxe^{i\theta})
\   f_\ell^{*}(k^{*}) + h_\ell^{(-)}
		 (kxe^{i\theta})\ f_\ell(k)\right ]\ .
\end{equation}
Note that the Jost functions in  Eqs.~(\ref{6}) and
(\ref{10}) are the same because  the Jost function does not depend on the
rotation angle $\theta$. Hence, the positions of its zeros are
$\theta$-independent as well.

To solve the differential equation, Eq.~(\ref{8}), we apply the variable
constant method\cite{vcm}, i.e. we look for a solution in the
form
\begin{equation}
\label{11}
    \Phi_\ell(k,xe^{i\theta})=\frac12\left [ h_\ell^{(+)}(kxe^{i\theta})
\    F_\ell^{(+)}(k,x,\theta) + h_\ell^{(-)}(kxe^{i\theta})
\    F_\ell^{(-)}(k,x,\theta)\right]\ ,
\end{equation}
which is an $r$--dependent combination
of  two linearly independent solutions of the free equation
\begin{equation}
\label{12}
    D_\ell(ke^{i\theta},x)\ h_\ell^{(\pm)}(kxe^{i\theta})=0.
\end{equation}
Instead of one unknown function $\Phi_\ell$, we have introduced two
unknown functions $F_\ell^{(\pm)}$ which necessitates
an additional constraint condition. The most convenient is the standard
Lagrange condition\cite{vcm}
\begin{equation}
\label{13}
   h_\ell^{(+)}(kxe^{i\theta})\ \partial_x F_\ell^{(+)}(k,x,\theta)+
   h_\ell^{(-)}(kxe^{i\theta})\ \partial_x F_\ell^{(-)}(k,x,\theta)=0\ ,
\end{equation}
which makes $\partial_x\,\Phi_\ell$ continuous even if the
potential has a sharp cut-off at a point $x=x_0$.

Substituting this ansatz, Eq.~(\ref{11}), into Eq.~(\ref{8}) and using
Eqs.~(\ref{12}) and (\ref{13}), lead to
coupled differential equations
\begin{eqnarray}
   \partial_x\,F_\ell^{(+)}(k,x,\theta) = &&\phantom{+} \frac{e^{i\theta}}
    {2ik}
    h_\ell^{(-)}(kxe^{i\theta})\ V(xe^{i\theta})\nonumber\\
 &&\times \left[h_\ell^{(+)}(kxe^{i\theta})\ F_\ell^{(+)}(k,x,\theta)
    + h_\ell^{(-)}(kxe^{i\theta})\ F_\ell^{(-)}(k,x,\theta)\right]
   \ ,
\label{14}
\end{eqnarray}
and
\begin{eqnarray}
  \partial_x\,F_\ell^{(-)}(k,x,\theta) = && -\frac{e^{i\theta}}{2ik}
  h_\ell^{(+)}(kxe^{i\theta})\ V(xe^{i\theta})\nonumber\\
  &&\times \left[h_\ell^{(+)}(kxe^{i\theta})\ F_\ell^{(+)}(k,x,\theta) +
   h_\ell^{(-)}(kxe^{i\theta})\ F_\ell^{(-)}(k,x,\theta)\right]\ ,
\label{14'}
\end{eqnarray}
with boundary conditions
\begin{equation}
\label{15}
       F_\ell^{(+)}(k,0,\theta)=F_\ell^{(-)}(k,0,\theta)=1\,,
\end{equation}
that follow immediately from Eqs.~(\ref{9}) and (\ref{11}).
With these boundary conditions, this system of coupled differential equations
are equivalent to the pair of integral Volterra-type equations,
\begin{equation}
    F_\ell^{(+)}(k,x,\theta)=1+\frac{e^{i\theta}}{ik}\int_0^x
    h_\ell^{(-)}(kx'e^{i\theta})V(x'e^{i\theta})\Phi_\ell(k,x'e^{i\theta})
\ dx'\ ,
\label{16}
\end{equation}
and
\begin{equation}
   F_\ell^{(-)}(k,x,\theta)=1-\frac{e^{i\theta}}{ik}\int_0^x
   h_\ell^{(+)}(kx'e^{i\theta})V(x'e^{i\theta})\Phi_\ell(k,x'e^{i\theta})
 \ dx'\ .
\label{16'}
\end{equation}
Iterated solution of these equations, on a small interval $[0,x_{min}]$
and with $\jmath_\ell(kxe^{i\theta})$ as the zero-order approximation to
$\Phi_\ell(k,xe^{i\theta})$, can be
used to obtain the boundary values $F_\ell^{(\pm)}(k,x_{min},\theta)$,
which in turn are
 needed to solve the differential equations, Eqs.~(\ref{14}) and (\ref{14'}),
numerically.
Thus, we can find the complex dilation $\Phi_\ell(k,xe^{i\theta})$ of
the regular solution in the form Eq.~(\ref{11}), using either  differential
(Eqs.~(\ref{14}) and~(\ref{14'})) or  integral (Eqs.~(\ref{16})
and~(\ref{16'})) equations.

 The question then arises as to the physical meaning of this
 solution and that  of the functions $F_\ell^{(\pm)}(k,x,\theta)$.
To answer, we note that the complex rotation, Eq.~(\ref{7}),
makes the Hamiltonian nonhermitian, so it can have complex eigenvalues
\cite{afnan}. Indeed, the Jost function can have zeros, not only
 on the imaginary axis of the (complex) $k$-plane but also in the entire
lower half of that plane. Schematically this is depicted in Fig.~1.
Therein, the points on the positive imaginary axis ($k_{01}, k_{02}, k_{03}$)
represent  Jost function zeros corresponding to physical bound states of
the system. The points shown in the fourth quadrant ($k_{04}, k_{05}$),
are the resonance state zeros.
The dividing line defined by the rotating angle $\theta$ is also displayed.
This line is the limit for the resonance position
angle $\varphi$ for which the condition $I\!m\, kr>0$ is fulfilled.

The asymptotic form of the function
$\Phi_\ell(k_0,xe^{i\theta})$, corresponding to these zeros,  decreases
exponentially if the angle $\theta$ is chosen large enough, i.e.
 greater than the angle $\varphi$ of Eq.~(\ref{7*}), since\cite{abram}
\begin{equation}
h_\ell^{(\pm)}(kxe^{i\theta})\mathop{\longrightarrow}\limits
_{x\to\infty}\mp i\exp[\pm i(kxe^{i\theta}-\frac{\ell\pi}{2})].
\end{equation}
Hence $\Phi_\ell(k_0,xe^{i\theta})$ are square-integrable eigenfunctions
of the rotated Hamiltonian  corresponding to either real (bound)
($R\!e\,k_0=0, I\!m\,k_0>0$)
or complex
($R\!e\,k_0>0, I\!m\,k_0<0$)
eigenvalues. Although these eigenvalues, being independent of $\theta$,
always correspond to the energies of the physical bound and resonance states,
 the corresponding solutions $\Phi_\ell(k_0,xe^{i\theta})$ acquire physical
meaning only when $\theta=0$. However, if one is interested only in
eigenenergies, it is  convenient to use $\theta>0$ thereby dealing
with square-integrable wave functions. This idea was successfully used for
the location of resonances with variational methods \cite{ccr}.

In the present approach, the square-integrable nature of the dilated
regular solution at  resonance energies is not used directly.
Instead, the crucial feature for our approach  is that the function
$h_\ell^{(+)}(kxe^{i\theta})$  decreases  exponentially  at
large $x$ if $I\!m\,(kxe^{i\theta})>0$. This is true for the whole region of
the complex $k$-plane in which the bound and resonance state zeros lie
(see Fig. 1). In contrast, the regular solution is square integrable
only at the discrete points $k_{0i}$ of this region. According to
Eq.~(\ref{10}), at all other points $k$ not
coinciding with the spectrum $k_{0i}$ of the rotated Hamiltonian,
the regular solutions diverge as
\begin{equation}
\Phi_\ell(k,xe^{i\theta})\sim e^{\displaystyle| I\!m\,
    (kxe^{i\theta})|}\ .
\end{equation}
As can be seen from Eqs.~(\ref{16}) and (\ref{16'}),
both functions $F_\ell^{(\pm)}$ at the spectral points $k_{0i}$,
have finite limits when $x\to\infty$
\begin{equation}
    \lim_{x\to\infty}F_\ell^{(+)}(k_{0i},x,\theta)=f_\ell^*(k_{0i}^*)\ ,
     \label{21}
\end{equation}
and
\begin{equation}
     \lim_{x\to\infty}F_\ell^{(-)}(k_{0i},x,\theta)=f_\ell(k_{0i})=0\ ,
\label{21'}
\end{equation}
which must coincide with the asymptotic coefficients defined in
Eq.~(\ref{10}). At any other point $k\ne k_{0i}$  within the zone
$I\!m\,(kxe^{i\theta})>0$, the integral of Eq.~(\ref{16'}) still has a finite
limit,
\begin{equation}
\label{19}
\lim_{x\to\infty}F_\ell^{(-)}(k,x,\theta)=f_\ell(k)\ ,
\end{equation}
because $h_\ell^{(+)}(kxe^{i\theta})$ is a decreasing function.
However, the  integral in Eq.~(\ref{16}) diverges.
 This is not surprising and by no means
a defect to our approach. Simply, it reflects the well-known
fact (see, for example, Refs.~\cite{newton} and \cite{gw}) that the  Jost
solutions, ${\cal F}_\ell
^{(\pm)}(k,r)$, of the Schr\"odinger equation with  boundary conditions,
\begin{equation}
       \lim_{r\to\infty}e^{\mp ikr}{\cal F}_\ell^{(\pm)}(k,r)=1\ ,
\end{equation}
are defined within different regions of the complex $k$-plane :
$I\!m\,k\le 0$ for ${\cal F}_\ell^{(+)}$ and
$I\!m\,k\ge 0$ for ${\cal F}_\ell^{(-)}$.
The only common region of their definition is the real axis $I\!m
\,k=0$   i.e. the physical region. Only for a special class of
rapidly decreasing potentials it would be a band along this axis
\cite{newton,taylor,gw}. However here we consider the
most general case defined by the condition Eq.~(\ref{2}).

The same is true for the limit ($x\to\infty$) of our functions
$F_\ell^{(\pm)}(k,x,\theta)$ with $\theta=0$, which are closely
related
to the Jost solutions, as one can see from Eq.~(\ref{11}). However, in
contrast
to the Jost solutions, the functions $F_\ell^{(\pm)}(k,x,0)$ are well
defined
for any finite $x$, because the boundary conditions, Eq.~(\ref{15}), are
imposed on them at the point $x=0$.

When $\theta>0$, the border dividing the two regions of complex $k$ where
the limits of $F_\ell^{(\pm)}(k,x,\theta)$  exist, is rotated downwards
to the line $|k| e^{-i\theta}$. The fact that above this line the function
$F_\ell^{(+)}(k,x,\theta)$ has no limit for $x\to\infty$ does not
contradict
the asymptotic requirement of Eq.~(\ref{10}).
Written in general form, these asymptotics must include both terms,
$$
h_\ell^{(+)}(kxe^{i\theta})f_\ell^{*}(k^{*})\qquad
{\rm and} \qquad
h_\ell^{(-)}(kxe^{i\theta})f_\ell(k),
$$
only on the dividing line, $I\!m\,(kxe^{i\theta})=0$,
where they are of the same order of magnitude. Above this line,
 Eq.~(\ref{10}) is a  sum of `small' and `large' terms; the first of
which can  be omitted. The `small' term,
$h_\ell^{(+)}(kxe^{i\theta})f_\ell^{*}(k^{*})$,
 becomes significant above the
dividing line only at the discrete points $k=k_{0i},i=1,2,\dots,N$ where
the `large' term, $h_\ell^{(-)}(kxe^{i\theta})f_\ell(k)$,
disappears because $f_\ell(k_{0i})=0$. And, as we
have already seen, at these points the limit Eq.~(\ref{21}) exists and the
`small' term is well defined. At all other points the growth of
$F_\ell^{(+)}(k,x,\theta)$ is suppressed by the exponentially decreasing
 factor $h_\ell^{(+)}(kxe^{i\theta})$ and the `small' term becomes
negligible as compared to the `large' term. Indeed, at a sufficiently large
$x$ the integrand of Eq.~(\ref{16}) behaves like
$$
\sim \exp(-2ikxe^{i\theta})V(xe^{i\theta}).
$$
If  the potential at infinity is a nonzero constant, then the integration
would produce $F_\ell^{(+)}\sim\exp(-2ikxe^{i\theta})$. However, the
potential
obeys the condition Eq.~(\ref{2}) so that $F_\ell^{(+)}$ diverges less rapidly
than $\exp(-2ipxe^{i\theta})$. Therefore the product
$h_\ell^{(+)}F_\ell^{(+)}$ grows less rapidly than
$h_\ell^{(-)}(kxe^{i\theta})$ and does not affect the asymptotic
properties specified  by
Eq.~(\ref{10}).
The two-term decomposition, Eq.~(\ref{11}), guarantees
term by term correspondence with the asymptotic form Eq.~(\ref{10}) at the
discrete points $k_{0i}$ and  also on the dividing line
$I\!m\,(kxe^{i\theta})=0$. For the other points
($k\ne k_{0i},I\!m\,(kxe^{i\theta})> 0$),
the first term of Eq.~(\ref{11}) contains some admixture $d_\ell(k,x,\theta)$
of the diverging term. However, this admixture  is infinitesimal as compared
to  the leading term,  $h_\ell^{(-)}(kxe^{i\theta})f_\ell(k)$
of the long range asymptotic behaviour,
\begin{equation}
   \Phi_\ell(k,x,\theta)\mathop{\longrightarrow}\limits_{x\to\infty}
   \frac12\left\{[h_\ell^{(+)}(kxe^{i\theta})f_\ell^{*}(k^{*})+
   d_\ell(k,x,\theta)]+h_\ell^{(-)}(kxe^{i\theta})f_\ell(k)\right\}\ ,
\label{lrasym1}
\end{equation}
i.e.
\begin{equation}
\lim_{x\to\infty}\left\{[h_\ell^{(+)}(kxe^{i\theta})f_\ell^{*}(k^{*})+
   d_\ell(k,x,\theta)]\left/h_\ell^{(-)}(kxe^{i\theta})f_\ell(k)\right.
   \right\}=0\ .
\label{divas}
\end{equation}
The asymptotic behaviour of the functions
$F_\ell^{(+)}(k,x,\theta)$ and $F_\ell^{(-)}(k,x,\theta)$ below the dividng
line $|k| e^{-i\theta}$ is just the opposite, i.e. $F_\ell^{(+)}$ is finite
and $F_\ell^{(-)}$ diverges. However, the function
$F_\ell^{(-)}(k,x,\theta)$ has a physical content only when
$I\!m\,(kxe^{i\theta})\ge 0$, while   $F_\ell^{(+)}(k_{0i},x,\theta)$
at the discrete points $k_{0i},i=1,2,
\dots,N$, where $F_\ell^{(-)}(k_{0i},\infty,\theta)=0$.

Thus, if the potential permits a large enough choice for $\theta$, we can
calculate the Jost function for all complex values of $k\ne 0$ of
interest, including the momenta on the positive imaginary
axis and in the resonance region under the positive real axis as
well.
As a consequence of
 Eq.~(\ref{19}), the Jost function can be found by solving
the coupled one--dimensional equations, Eqs.~(\ref{14}) and~(\ref{14'}),
from $x=0$ to $x=x_{max}$, where $x_{max}$ is
determined by the required accuracy  and on how fast the potential decreases.
 The faster it decreases the sooner
$\partial_x\,F_\ell^{(-)}(k,x,\theta)$ vanishes and
$F_\ell^{(-)}(k,x,\theta)$ tends smoothly to the constant $f_\ell(k)$.

It is interesting to note that at sufficiently
large $x$ the system of equations, Eqs.~(\ref{14}) and~(\ref{14'}),
decouple as
the `small' term
 $h_\ell^{(+)}F_\ell^{(+)}$ becomes insignificant as compared to the
`large' term $h_\ell^{(-)}F_\ell^{(-)}$.

\section{Noncentral short--range potential}

In this section we consider a two-body problem with spin-dependent
interaction which does not
conserve the orbital angular momentum $\ell$. As in the previous
section, we assume that the matrix elements
$$
V_{(JM\ell's'),(JM\ell s)}^{JM}(r)\equiv <{\cal Y}_{\ell's'}^{JM}|
V(\vec r)| {\cal Y}_{\ell s}^{JM}>
$$
of the potential operator, sandwitched between the spin-angular momentum
eigenfunctions
$$
{\cal Y}_{\ell s}^{JM}({\bf \hat  r})\equiv \sum_{m\mu}\,C_{\ell m
s\mu}^{JM} Y_{\ell m}({\bf \hat  r})\chi_{s\mu}
$$
of the total angular momentum operator,
$$
\vec J=\vec\ell+\vec s  ,
$$
obey the same constraints (\ref{1}) and (\ref{2}) at the origin and infinity.

To simplify the notation, the superscripts $JM$ will be omitted and instead
of the pair of subscripts $\ell s$, satisfying the triangle condition
$$
|J-s|\le \ell\le |J+s| ,
$$
we will use the symbol $[\ell]$.

\subsection{Boundary conditions}
The physical wave-function, describing the state with a given momentum
$\vec k$, is the product
$$
       \Psi(\vec k,\vec r)=\frac{1}{r}({\cal Y}\cdot u)\equiv\frac{1}{r}
       \sum_{[\ell]}\,{\cal Y}_{\ell s}^{JM}(\hat{{\bf r}})u_{[\ell]}
       (\vec k,r)
$$
of the $[\ell]$--row of { ${\cal Y}$ and the  $[\ell]$--column $u$ of the
radial components   satisfying the coupled differential equations
\begin{equation}
\label{fiz}
      D_\ell(k,r)u_{[\ell]}
      (\vec k,r)=\sum_{[\ell']}V_{[\ell][\ell']}(r)
      u_{[\ell']}(\vec k,r)
 \end{equation}
with  (physical) boundary conditions
\begin{eqnarray}
      && u_{[\ell]}(\vec k,0)=0,\nonumber\\
       \phantom{-}\label{inf}\\
      && u_{[\ell]}(\vec k,r)\mathop{\longrightarrow}\limits_{r\to\infty}
       U_{[\ell]}(\vec k,r),\nonumber
\end{eqnarray}
where the asymptotic function $U_{[\ell]}(\vec k,r)$ has a
different form for the bound, scattering, and resonance states.

As it is more convenient to deal with  universal boundary
conditions imposed at a single point, we consider the
so-called fundamental matrix, $\Phi_{[\ell'][\ell]}(k,r)$, of the regular
solutions instead of the physical column solution, $u_{[\ell]}(\vec k,r)$.
The physical solution then can be constructed as a linear combination of
columns of $\Phi_{[\ell'][\ell]}(k,r)$.
By  definition, $\Phi$ is a fundamental matrix if it has the
following three properties \cite{newton}:\\
\indent i) Each column of $\Phi_{[\ell'][\ell]}$ is a solution of
      Eq.~(\ref{fiz}), i.e
\begin{equation}
 \label{regeq}
     D_\ell(k,r)\ \Phi_{[\ell][\ell']}(k,r)=\sum_{[\ell'']}
     V_{[\ell][\ell'']}(r)\ \Phi_{[\ell''][\ell']}(k,r)\ ,
\end{equation}
\indent ii) $\Phi_{[\ell'][\ell]}(k,0)=0$,\quad and\\
\vskip 0.1cm
\indent iii) all columns of $\Phi_{[\ell'][\ell]}$ are linearly
    independent.\\
Moreover, $\Phi_{[\ell'][\ell]}$ must be a square matrix since
Eq.~(\ref{fiz}) has as many independent regular solutions as the column
dimension \cite{kamke}.\\
The property ii) implies that the desired universal boundary condition
should be imposed at the point $r=0$.
We emphasize here that the boundary condition
$\Phi_{[\ell'][\ell]}(k,0)=0$, is not sufficient to
define this matrix since Eq.~(\ref{regeq}) is of second order. Furthermore,
Eq.~(\ref{16}) is singular at the origin, and
consequently the existence and uniqueness theorem\cite{kamke}
is not valid at this point. However, this theorem is valid for
any small $r \ge \delta > 0$.
Hence, the matrix $\Phi$ can be uniquely defined by
$\Phi(k,\delta)$ and $\partial_r\Phi(k,\delta)$, while  within the
interval $[0,\delta]$,  $\Phi(k,r)$ can be evaluated explicitly
as follows.
After multiplying by $r^2$ and using Eq.~(\ref{1}), Eqs.~(\ref{regeq})
 decouple  giving
\begin{equation}
\left[r^2\partial_r^2-\ell(\ell+1)\right]\Phi_{[\ell][\ell']}(k,r)=0,
\quad r\in[0,\delta]\ .
\end{equation}
Each of these equations has two independent
solutions, namely the functions $\sim r^{\ell+1}$ and $\sim r^{-\ell}$.
Therefore, near the origin, we can use the Riccati-Bessel function,
\begin{equation}
\jmath_\ell(kr)\,\mathop{\longrightarrow}\limits_{r\to0}\,\,
\displaystyle{\frac{\sqrt{\pi}k^{\ell+1}}{2^{\ell+1}\Gamma(\ell+
\frac{3}{2})}}r^{\ell+1}\ ,
\end{equation}
as the regular solution.
 Hence, as a first order approximation to $\Phi_{[\ell][\ell']}(k,r)$
for $r\in[0,\delta]$ we can take  the diagonal matrix
\begin{equation}
\Phi_{[\ell][\ell']}(k,r)\mathop{\sim}\limits_{r\to 0}
\delta_{[\ell][\ell']}\ \jmath_{\ell'}(kr)\ ,
\end{equation}
satisfying the above-mentioned three properties.
Higher order corrections give non--zero off-diagonal elements, so
to guarantee linear independence of the columns,
the off-diagonal elements of a row must be infinitesimal in comparison to
the diagonal element of that row. The boundary condition  thus reads
\begin{equation}
\label{bc}
     \lim_{r\to 0}\displaystyle{\frac{\Phi_{[\ell][\ell']}(k,r)}
     {\jmath_{\ell'}(kr)}}=\delta_{[\ell][\ell']}\ ,
\end{equation}
which is a generalization of Eq.~(\ref{5}).

Since $\Phi$ is the fundamental matrix, any physical solution is  a linear
combination of its columns with $r$-independent coefficients, i.e.
\begin{equation}
\label{physcon}
  u_{[\ell]}(\vec k,r)
     =\sum_{[\ell']}\Phi_{[\ell][\ell']}(k,r)A_{[\ell']}(\vec k)\ .
\end{equation}
The coefficients $A_{[\ell]}$ must be  chosen  to
satisfy the appropriate boundary condition (\ref{inf}) at infinity.
 \subsection{Complex rotation}
To generalize the approach described in the previous
section, we perform the complex
rotation, Eq.~(\ref{7}), of the coordinate. The corresponding
complex dilation of the fundamental
matrix, $\Phi_{[\ell][\ell']}(k,xe^{i\theta})$, is the solution of
\begin{equation}
      D_\ell(ke^{i\theta},x)\ \Phi_{[\ell][\ell']}(k,xe^{i\theta})=
      e^{2i\theta}\sum_{[\ell'']}V_{[\ell][\ell'']}(xe^{i\theta})
      \ \Phi_{[\ell''][\ell']}(k,xe^{i\theta})\ ,
\end{equation}
where
\begin{equation}
\label{a}
     \lim_{x\to0}\left [\Phi_{[\ell][\ell']}(kxe^{i\theta})
    \left/j_{\ell'}(kxe^{i\theta})\right.\right
	  ]=\delta_{[\ell][\ell']}\ .
\end{equation}
We then apply  the variable-constant method i.e. look for a solution of the
form
\begin{equation}
\label{b}
       \Phi_{[\ell][\ell']}(k,xe^{i\theta})=\frac12 \left[
       h_\ell^{(+)}(kxe^{i\theta})
      \  F_{[\ell][\ell']}^{(+)}(k,x,\theta)+h_\ell^{(-)}(kxe^{i\theta})
      \  F_{[\ell][\ell']}^{(-)}(k,x,\theta)\right]\ ,
\end{equation}
with
\begin{equation}
       h_\ell^{(+)}(kxe^{i\theta})\,\partial_x
       F_{[\ell][\ell']}^{(+)}(k,x,\theta)+
       h_\ell^{(-)}(kxe^{i\theta})\,\partial_x
      F_{[\ell][\ell']}^{(-)}(k,x,\theta)=0\ .
\end{equation}
The resulting coupled equations  for the new unknown  functions,
 $F_{[\ell][\ell']}^{(\pm)}(k,x,\theta)$, have the form
\begin{eqnarray}
      \partial_x\,F_{[\ell][\ell']}^{(+)}(k,x,\theta)=
		&&\phantom{+} \frac{e^{i\theta}}{2ik}
       h_\ell^{(-)}(kxe^{i\theta})\sum_{[\ell'']}
		   V_{[\ell][\ell'']}(xe^{i\theta})
\nonumber\\
    && \times \left[ h_{\ell''}^{(+)}(kxe^{i\theta})
		  F_{[\ell''][\ell']}^{(+)}(k,x,\theta)+
	 h_{\ell''}^{(-)}(kxe^{i\theta})F_{[\ell'']
		[\ell']}^{(-)}(k,x,\theta)\right]\ ,
\label{c}
\end{eqnarray}
and
\begin{eqnarray}
\partial_x\,F_{[\ell][\ell']}^{(-)}(k,x,
		 \theta)= && -\frac{e^{i\theta}}{2ik}
       h_\ell^{(+)}(kxe^{i\theta})\sum_{[\ell'']}
		  V_{[\ell][\ell'']}(xe^{i\theta})
\nonumber\\
       &&\times \left [ h_{\ell''}^{(+)}(kxe^{i\theta})
		   F_{[\ell''][\ell']}^{(+)}(k,x,\theta)+
  h_{\ell''}^{(-)}(kxe^{i\theta})
		  F_{[\ell''][\ell']}^{(-)}(k,x,\theta)\right]\ .
\label{c'}
\end{eqnarray}
The appropriate boundary conditions for them can be derived from
Eqs.~(\ref{a}) and~(\ref{b}).
First though, one must compensate  the singularities of
$h_\ell^{(+)}(kxe^{i\theta})$ and $h_\ell^{(-)}(kxe^{i\theta})$ at $x=0$.
For this we can demand
$F^{(+)}$ and $F^{(-)}$ to be identical as $x\to 0$, viz.
\begin{equation}
     F_{[\ell][\ell']}^{(+)}(k,x,\theta)\mathop{\sim}\limits_{x\to0}
     F_{[\ell][\ell']}(k,x,\theta)\ ,
\end{equation}
and
\begin{equation}
     F_{[\ell][\ell']}^{(-)}(k,x,\theta)\mathop{\sim}\limits_{x\to0}
     F_{[\ell][\ell']}(k,x,\theta)\ ,
\end{equation}
as then
\begin{equation}
\label{e}
      \Phi_{[\ell][\ell']}(k,xe^{i\theta})\mathop{\sim}\limits_{x\to0}
      \jmath_\ell(kxe^{i\theta})\ F_{[\ell][\ell']}(k,x,\theta)\ ,
\end{equation}
so that, from Eq.~(\ref{a}), the boundary conditions become
\begin{equation}
\label{d}
     \lim\limits_{x\to0}\left[\jmath_\ell(kxe^{i\theta})
    F_{[\ell][\ell']}^{(\pm)}(k,x,\theta)
        { \left/\jmath_{\ell'}(kxe^{i\theta}) \right.}
         \right] =\delta_{[\ell][\ell']}\ .
\end{equation}
However, as $x=0$ is a singular point, for
practical calculations one needs to solve Eqs.~(\ref{c}) and (\ref{c'})
analytically on a small interval $(0,\delta]$ and then  impose the
boundary conditions at $x=\delta$. Such an analytical solution can be found
easily by choosing $\delta$ to be small enough so that for $x\in(0,\delta]$
 we can write
\begin{equation}
\partial_xF_{[\ell][\ell']}^{(\pm)}(k,x,\theta)=\pm\frac{e^{i\theta}}{ik}
h_\ell^{(\mp)}(kxe^{i\theta})\ V_{[\ell][\ell']}(xe^{i\theta})
\ \jmath_{\ell'}(kxe^{i\theta})\ .
\end{equation}
For small $x$, in $h_\ell^{(\pm)}\equiv\jmath_\ell\pm in_\ell$,
the Riccati-Neumann function $n_\ell$ is dominant, so that
\begin{equation}
\partial_xF_{[\ell][\ell']}^{(\pm)}(k,x,\theta) \sim -\frac{e^{i\theta}}{k}
n_\ell(kxe^{i\theta})\ V_{[\ell][\ell']}(xe^{i\theta})\ \jmath_{\ell'}
(kxe^{i\theta})\ .
\end{equation}
Upon integration of this (approximate) equation one finds,
\begin{equation}
F_{[\ell][\ell']}^{(\pm)}(k,x,\theta) = -\frac{e^{i\theta}}{k}\int
n_\ell(kxe^{i\theta})\ V_{[\ell][\ell']}(xe^{i\theta})\ \jmath_{\ell'}
(kxe^{i\theta})\
d\!x
+{\rm const}\ ,
\end{equation}
and if the arbitrary constant of integration is taken to be equal to
$\delta_{[\ell][\ell']}$, this function has the asymptotic behaviour,
\begin{equation}
\label{f}
F_{[\ell][\ell']}^{(\pm)}(k,x,\theta)\mathop{\sim}\limits_{x\to0}
\delta_{[\ell][\ell']}-\frac{e^{i\theta}}{k}\int n_\ell(kxe^{i\theta})
\ V_{[\ell][\ell']}(xe^{i\theta})\ \jmath_{\ell'}(kxe^{i\theta})\ d\!x\ ,
\end{equation}
which obeys the condition of Eq.~(\ref{d}). In practical calculations the
last indefinite integral can be found analytically by using the leading
terms of series expansions of $n_\ell,\,V_{[\ell][\ell']},\,{\rm and}\,
\jmath_{\ell'}$.
In this way one can find that the second term of Eq.~(\ref{f}) is regular at
$x=0$ if $\ell\le\ell'$ (right upper corner and the diagonal of the matrix)
and may be singular if $\ell>\ell'$ (left lower corner of the matrix).
But this singularity is compensated by the presence of
$\jmath_\ell$ in Eq.~(\ref{e}).

Thus, the coupled equations, Eqs.~(\ref{c}) and (\ref{c'}),
along with the boundary conditions, Eqs.~(\ref{d}) and (\ref{f}),
form a well-defined differential problem.

By analogy with the case of a central potential, we find that
$F_{[\ell][\ell']}^{(+)}(k,x,\theta)$ has a
finite $\theta$-independent limit,
\begin{equation}
\lim_{x\to\infty}F_{[\ell][\ell']}^{(+)}(k,x,\theta)=
f_{[\ell][\ell']}^{(+)}(k)\ ,
\end{equation}
when $I\!m\, (kxe^{i\theta})\le 0$ (i.e. in the
complex $k$--plane below the dividing line), while
$F_{[\ell][\ell']}^{(-)}$ has the limit,
\begin{equation}
\lim_{x\to\infty}F_{[\ell][\ell']}^{(-)}(k,x,\theta)=
f_{[\ell][\ell']}^{(-)}(k)\ ,
\end{equation}
when $I\!m\,(kxe^{i\theta})\ge 0$ (i.e.
in the complex $k$--plane above that line).
Therefore, the  fundamental matrix has the asymptotic form,
\begin{equation}
\label{matas}
     \Phi_{[\ell][\ell']}(k,xe^{i\theta}) \mathop{\longrightarrow}
     \limits_{x\to\infty}\frac12\left[ h_\ell^{(+)}(kxe^{i\theta})
    \ f_{[\ell][\ell']}^{(+)}(k)+h_\ell^{(-)}
     (kxe^{i\theta})\ f_{[\ell][\ell']}^{(-)}(k)\right],
\end{equation}
with  $f_{[\ell][\ell']}^{(-)}(k)$ being the matrix generalisation of
the Jost function. We  refer to this as the {\it Jost matrix}.

Substituting Eq.~(\ref{matas}) into Eq.~(\ref{physcon}), one finds that the
physical solutions involve only the first term of Eq.~(\ref{matas}) if
\begin{equation}
\label{homo}
\sum_{[\ell']}f_{[\ell][\ell']}^{(-)}(k)A_{[\ell']}(\vec k)=0\,.
\end{equation}
This system of homogeneous algebraic equations defining the
coefficients $A_{[\ell]}$ has a nontrivial solution if and only if
\begin{equation}
\det\Vert f_{[\ell][\ell']}^{(-)}(k_0)\Vert =0\ .
\end{equation}
Therefore the zeros $k_{0i}$ of the determinant of the Jost matrix  are the
spectral points corresponding to the bound and resonance states.
The square-integrable complex-dilated wave function of any such  state is
\begin{eqnarray}
\Psi(k_0,\vec r,\theta)=\frac{e^{-i\theta}}{2x}\sum_{[\ell][\ell']}
{\cal Y}_{\ell s}^{JM}({\bf \hat r})
\ \left[ h_{\ell}^{(+)}(k_0xe^{i\theta})\ F_{[\ell][\ell']}^{(+)}(k_0,x,
\theta)\right.\nonumber\\
\left. + h_{\ell}^{(-)}(k_0xe^{i\theta})\ F_{[\ell][\ell']}^{(-)}(k_0,x,
\theta)\right]\ A_{[\ell']}(k_0)\ .
\end{eqnarray}

\section{Potential with Coulombic tail}

Consider the case  where the leading term of the long--range behaviour of the
potential has the form $2\eta k/r$. We split the potential as
\begin{equation}
V(r) = \{ V(r) -  { {2k\eta}\over r}\}
+ { {2k\eta}\over r}
\end{equation}
so that the entry in brackets is a short--range interaction. We designate
that hereafter as $V_{sr}(r)$ and note that it
 satisfies the constraints Eqs.~(\ref{1}) and (\ref{2}).
If the radial Schr\"odinger equation now is written as
\begin{equation}
D_\ell^\eta(k,r)\Phi_\ell(k,r)=V_{sr}(r)\Phi_\ell(k,r)\ ,
\end{equation}
where
\begin{equation}
D_\ell^\eta(k,r)\equiv\partial_r^2+k^2-\ell(\ell+1)/r^2-2\eta k/r\ ,
\end{equation}
the natural generalisation of the boundary condition, Eq.~(\ref{5}), for the
regular solution is
\begin{equation}
\lim_{r\to0}[\Phi_\ell(k,r)/F_\ell(\eta,kr)] = 1\ ,
\end{equation}
with $F_\ell(\eta,kr)$ being the standard  regular solution of the
Coulomb equation\cite{abram},
\begin{equation}
D_\ell^\eta(k,r)F_\ell(\eta,kr)=0\ .
\end{equation}
Application of the complex rotation, Eq.~(\ref{7}), gives
\begin{equation}
D_\ell^\eta(ke^{i\theta},x)\ \Phi_\ell(k,xe^{i\theta}) =
e^{2i\theta}V_{sr}(xe^{i\theta})\ \Phi_\ell(k,xe^{i\theta})\ ,
\end{equation}
and
\begin{equation}
\lim_{x\to0}\left[\Phi_\ell(k,xe^{i\theta})\left/
F_\ell(\eta,kxe^{i\theta})
\right.\right]=1\ .
\end{equation}
Then, following the procedure developed in Sec.~II,
solutions $\Phi_\ell(k,xe^{i\theta})$, are sought that have the form
\begin{equation}
\Phi_\ell(k,xe^{i\theta})=\frac12\left[ H_\ell^{(+)}(\eta,kxe^{i\theta})
\ F_\ell^{\eta(+)}(k,x,\theta)+H_\ell^{(-)}(\eta,kxe^{i\theta})
\ F_\ell^{\eta(-)}(k,x,\theta)\right]\ ,
\end{equation}
where now the Lagrange condition is
\begin{equation}
H_\ell^{(+)}(\eta,kxe^{i\theta})\,\partial_x F_\ell^{\eta(+)}(k,x,\theta)+
H_\ell^{(-)}(\eta,kxe^{i\theta})\,\partial_x F_\ell^{\eta(-)}(k,x,\theta)
= 0\ .
\end{equation}
Therein $H_\ell^{(\pm)}$ are the combinations of the regular and irregular
Coulomb functions,
\begin{equation}
H_\ell^{(\pm)}(\eta,z)\equiv F_\ell(\eta,z)\mp iG_\ell(\eta,z)\ ,
\end{equation}
having  the asymptotic behaviour,
\begin{equation}
H_\ell^{(\pm)}(\eta,z)\mathop{\longrightarrow}\limits_{|z|\to\infty}
\mp i \exp\Bigl\{\pm i\{z-\eta\ln 2z-\frac{\ell\pi}{2}+{\rm arg}\Gamma(\ell+
1+i\eta)\Bigr\}\ .
\end{equation}
In the limit $\eta \to 0$, these are reduced to the Riccati--Hankel functions
$h_\ell^{(\pm)}(z)$.

The above process  gives first order coupled equations for the
$F_\ell^{\eta(\pm)}(k,x,\theta)$, namely
\begin{eqnarray}
     \partial_x\,F_\ell^{\eta(+)}(k,x,\theta) &=&
     \phantom{+}\frac{e^{i\theta}}{2ik}
     H_\ell^{(-)}(\eta,kxe^{i\theta})\ V_{sr}(xe^{i\theta})\nonumber\\
  && \times \left[H_\ell^{(+)}(\eta,kxe^{i\theta})\ F_\ell^{\eta(+)}(k,x,
     \theta)+
     H_\ell^{(-)}(\eta,kxe^{i\theta})\ F_\ell^{\eta(-)}(k,x,\theta)\right]\ ,
\end{eqnarray}
and
\begin{eqnarray}
     \partial_x\,F_\ell^{\eta(-)}(k,x,\theta) &=& -\frac{e^{i\theta}}{2ik}
     H_\ell^{(+)}(\eta,kxe^{i\theta})\ V_{sr}(xe^{i\theta})\nonumber\\
   && \times \left[H_\ell^{(+)}(\eta,kxe^{i\theta})\ F_\ell^{\eta(+)}(k,x,
     \theta)+
    H_\ell^{(-)}(\eta,kxe^{i\theta})\ F_\ell^{\eta(-)}(k,x,\theta)\right]\ .
\end{eqnarray}
It is easily seen that the asymptotic properties of
$F_\ell^{\eta(\pm)}(k,x,\theta)$ are the same as those of
$F_\ell^{(\pm)}(k,x,\theta)$
discussed in Sec. II. Therefore the limit
\begin{equation}
\lim_{x\to\infty}F_\ell^{\eta(-)}(k,x,\theta)=f_\ell^\eta(k)\ ,
\end{equation}
exists for all complex $k$ obeying the condition $I\!m\,(kxe^{i\theta})
\ge0$  (i.e. above the dividing line).
The Coulomb modified Jost function is thus $f_\ell^\eta(k)$, and it follows
that the corresponding long range limit of $F_\ell^{\eta(+)}(k,x,\theta)$
exists in the zone $I\!m\,(kxe^{i\theta})\le0$ and for all
spectral points $k_{0i}$ where
$f_\ell^\eta(k_{0i})=0$, \,$i=1,2,\dots$ .

\section{A numerical example}

To illustrate  the method, we consider $s$--wave solutions for
the exponential potential,
\begin{equation}
\label{expot}
         V(r)=-\frac{2m}{\hbar^2}v\exp(-r/R)\ .
\end{equation}
For this case, the differential problem, defined by Eqs.~(\ref{3}) and
(\ref{5}), can be solved analytically\cite{newton}. The corresponding Jost
function is given by
\begin{equation}
\label{jex}
      f_0(k) = \Gamma(1-2ikR)\
           J_{-2ikR}\left(2R\sqrt{{2mv}\over \hbar^2}\right)
      \left(R\sqrt{{2mv}\over \hbar^2}\right)^{2ikR}\ ,
\end{equation}
and is defined for all complex $k$ except at the discrete
points $k_n=-in/(2R),\, n=1,2,3,\dots$ on the imaginary axis.

There are two parameter sets for this potential that are of particular
interest. The first, with  strength  $v=155.17\,{\rm MeV}$ and  range
$R=0.76\,{\rm fm}$, describes the triplet  $NN$-interaction \cite{brown}
( $\hbar^2/(2\mu)$=41.47 MeV--fm$^2$).
With this interaction, the Jost function has a zero on the positive
imaginary axis in the complex $k$--plane, corresponding to the  deuteron
 binding energy. The second parametrisation, is  $v=104.2\,{\rm MeV}$ and
$R=0.73\,{\rm fm}$ for the singlet $NN$-interaction \cite{brown}.
In this case there is a zero of the Jost function on the negative
imaginary axis, corresponding to  virtual singlet deuteron.
More specifically, these zeros of the function (\ref{jex}) lie at
 $k= i\ 0.23640\,{\rm fm}^{-1}\,\,(-2.3175\,{\rm MeV})$
and at $k= -i\ 0.039913\,{\rm fm}^{-1}\,\, (-0.066064\,{\rm MeV})$
respectively.
The relevant solutions, $F_0^{(-)}(k,x_{max},\theta)$, of Eqs.~(\ref{14})
and~(\ref{14'}), obtained by means of numerical Runge-Kutta integration up to
$x_{max}=20\,{\rm fm}$, have zeros at the same  momenta for all choices
of $\theta\in[0,\frac{\pi}{2})$.

The successful evaluation of the virtual state via
$F_0^{(-)}(k,x_{max},\theta)$, despite the fact that the corresponding
 zero lies below the dividing line, is due  to the
short range nature of the potential. In fact,
for such potentials the non-dilated Jost solutions are both well defined
within a narrow band along the dividing line\cite{newton}; a band
 wide enough to include the zero linked to the virtual state.

To check the equivalence of $F_0^{(-)}(k,x_{max},\theta)$ to the exact
Jost function  for different values of complex momenta and
to observe $\theta$-independence of $F_0^{(-)}(k,x_{max},\theta)$ above
the dividing line, we used the triplet deuteron potential (first set of
parameter values) and obtained the  solutions for six values of the momenta
that are  situated on the arc of the radius $ | k |=2\,{\rm fm}^{-1}$.
With $k =\vert k\vert \exp(i\varphi_n)$, we chose
$ \varphi_n$ to be $\pi/2$, $\pi/4$, 0, $-\pi/8$, $-\pi/4$, and $-3\pi/8$.
The  calculations were repeated with  four different
 values of the dilation angle $\theta$ (0, $\pi/8$, $\pi/4$, and  $3\pi/8$ )
which correspond to different directions of the dividing line (see Fig.~2).
A point $| k | \exp(i\varphi_n)$ is above this line when $I\!m\,(kxe^
{i\theta})\ge0$, i.e. when $\varphi_n+\theta\ge0$.

The results are presented in Table I.
For non-negative values of $\phi$, the calculated functions agree
very well (to 1 part in 10$^5$) with the exact Jost function
 and the solutions are very evidently $\theta$--independent.
For negative $\phi$ however, close agreement with the exact Jost function
values and $\theta$--independence of the calculations only
occurs with the choice $\theta \ge \vert\varphi\vert$.
As expected from our theory, the divergence is extreme when this condition
is violated.

These results  demonstrate the effect of the complex rotation in that
the greater one can take $\theta$, the wider is the sector of the
fourth quadrant of complex $k$-plane  where $F_0^{(-)}(k,x_{max},\theta)$
coincides with the (exact in the test case) Jost function
and therein the calculated function is independent of $\theta$.

In Table~II, the results of our calculations of the test example are shown
for different coordinate `$x$' values in the range 0.5 $\le x \le$ 20\,fm
 but at a fixed momentum and with $\theta$ set to zero.
These results clearly display the divergence of $F_0^{(+)}(k,x,0)$ and
the convergence of $F_0^{(-)}(k,x,0)$ when $x\to\infty$.
The third column of this table shows a rapid convergence of
$F_0^{(-)}$ which reflects  the fact that this function is
smoother than the whole function $\Phi_0$ and varies only at
distances where the potential is significant.  Indeed, $F_0^{(-)}$ has
converged very well (~1~part in 10$^5$) by $x = 10$. But the relevant
asymptotic terms are the ratios of the terms in Eq.~(\ref{11}).
As $x\to \infty$, these ratios decrease rapidly.  The last column of Table~II
explicitly shows how the first (`small') term, $h_0^{(+)}F_0^{(+)}$,
of the asymptotics of Eq.~(\ref{divas}), becomes infinitesimal in  comparison
with the second (`large') term, $h_0^{(-)}F_0^{(-)}$, despite the divergence
of $F_0^{(+)}(k,x,0)$.

\section{Conclusions}

We have shown how a combination of the variable--constant and complex
coordinate rotation  methods can be used to recast the
two--body Schr\"{o}dinger equation with the diverse boundary conditions to
consider bound, virtual, scattering, and resonance states,
into a set of linear first order coupled equations for  Jost--type solutions,
 with which all possible state forms can be treated in a unified way.
The derived equations are amenable to numerical intergration
to obtain the Jost functions (Jost matrices) for all momenta of physical
 interest. The viability of this approach was demonstrated  by its
application to find solutions for a potential for which the ($s$--wave)
 Jost function has an analytic form.

An advantage of the proposed systems of equations,  over studies of
two--body states in which the Schr\"odinger or  Lippmann-Schwinger equations
are used, is that the bound state and scattering problems
can be treated by the same method. In addition, resonance states can also be
handled. Moreover, for the scattering
problem a much higher accuracy can be achieved because both $F^{(+)}$ and
$F^{(-)}$ are smooth functions since the oscillatory behaviour of the
asymptotic form of the complete solution $\Phi$ is factored out in
 this approach.

 With our approach, one can examine the contribution of
different parts of the potential curve in the scattering process,
 simply by truncating the integrations at different distances.
That is equivalent to truncation of the potential.  This
information is contained in $F^{(-)}$ (~ as a function of $r$ ) because it
is equal to the Jost function for the truncated potential at distance $r$.

Another important advantage of our approach over the conventional ones,
is that one can calculate the Jost function for complex momenta in the
 resonance region.  To our knowledge, this is the simplest way to locate
 resonances (zeros of $\det F^{(-)}(k,x_{max},\theta)$), with high accuracy.
Such information is of crucial importance in various fields of Physics.
 Inverse scattering problems where the parametrization of the $S$ matrix
 is required, is a case. In the $\ell$-dependent Marchenko inversion method
\cite{newton}, for example, the $S$ matrix is written as
\begin{equation}
S_\ell(k)=\frac{f_\ell(-k)}{f_\ell(k)}=\prod_n\frac{k+\alpha_n}{k-\alpha_n},
\end{equation}
and the interpretation of the parameters and their location in the
complex $k$-plane are important in order to obtain the correct potential.

The proposed method has certain drawbacks however. One  is that when
$\theta\ne 0$, analytic continuation of the potential to the first quadrant
of the complex $r$-plane is required. That may not be easy to do
 when the potential function of interest has discontinuities
or is given in the form of a numerical table.
A second drawback concerns the point $k=0$ at which the proposed
 equations are singular. This problem can be overcome
by using a prescription given in Ref.~\cite{zero}. Within a small region
around $k=0$, the Riccati-Hankel functions $h_\ell^{(\pm)}(kr)$ can
be expanded in a power series\cite{abram}  and each term therein factorized
 in $k$ and $r$. Similarly  $F^{(\pm)}$ can be expanded in powers of $k$
with unknown $r$-dependent coefficients in this region; the coefficients
being specified by the resulting system of coupled, $k$-independent,
differential equations.

The range of application of the proposed  method can be extended further.
Since the derived formulae are valid for non--integer values of the angular
 momentum $\ell$, the method can be used to solve the $N$-body hyperradial
 equation of the hyperspherical harmonics approach \cite{fabr} and which
 has the form of  the radial Schr\"odinger equation but with half integer
values for $\ell$. Using non--integer $\ell$, we can also consider
 potentials that tend as $\sim r^{-2}$ toward the origin, by including
the singularity of the potential into the centrifugal term.
 Such cases arise when  supersymmetric  (SUSY)
transformations are applied to remove a bound state from the spectrum.
The resulting supersymmetric partner potential has a $\sim 1/r^2$ behaviour
 at the origin\cite{baye}.

Finally, we can consider complex values of the angular momentum $\ell$,
and, in principle, study all the problems concerning Regge trajectories.
\bigskip
\begin{center}
{\large ACKNOWLEDGEMENTS}
\end{center}
Financial support from the  University of South
Africa  and the Joint Institute  for Nuclear Research, Dubna,
given to S.A.R and S.A.S. is greatly appreciated.
Likewise, K.A. gratefully acknowledges the financial support
of the F.R.D. of South Africa that made possible a visit
to Pretoria during which this work was initiated.


\newpage


\begin{table}
\caption[]
 {Comparison of $F_0^{(-)}(k,{\rm x =}\,20\,{\rm fm.}, \theta)$ with the
exact Jost function (Eq.~(\ref{jex})) at  the six momentum points situated
on the arc defined by $k=2\ e^{i\varphi}\ {\rm fm}^{-1}$.
The numbers given in the brackets give the exponent power of ten.}
\vskip 0.2cm
\begin{tabular}{c@{\hspace{8mm}}c@{\hspace{8mm}}c@{\hspace{8mm}}c@
	 {\hspace{8mm}}c@{\hspace{8mm}}c@{\hspace{8mm}}c}

$\varphi$ & $f_0(p)$ & $\theta=0$ & ${\displaystyle\theta=\frac{\pi}{8}}$ &
${\displaystyle\theta=\frac{\pi}{4}}$ &
$\displaystyle{\theta=\frac{3^{\mathstrut}\pi}{8_{\mathstrut}}}$ & exact \\
\tableline
${\displaystyle\frac{\pi}{2}}$ & $\displaystyle{R^{\mathstrut}e\atop
I_{\mathstrut}m}$ &
$\displaystyle{0.56706\atop 0.}$ &
$\displaystyle{0.56706\atop -0.32\,\ (-8)}$ &
$\displaystyle{0.56706\atop0.74\,\ (-9)}$ &
$\displaystyle{0.56706\atop0.12\,\ (-4)}$ &
$\displaystyle{0.56706\atop0.}$ \\\\

${\displaystyle\frac{\pi}{4}}$ & $\displaystyle{R^{\mathstrut}e\atop
I_{\mathstrut}m}$ &
$\displaystyle{0.59205\atop -0.21965}$ &
$\displaystyle{0.59205\atop -0.21965}$ &
$\displaystyle{0.59205\atop -0.21965}$ &
$\displaystyle{0.59205\atop -0.21963}$ &
$\displaystyle{0.59205\atop -0.21965}$ \\\\

$0$ & $\displaystyle{R^{\mathstrut}e\atop I_{\mathstrut}m}$ &
$\displaystyle{0.69491\atop -0.48975}$ &
$\displaystyle{0.69491\atop -0.48975}$ &
$\displaystyle{0.69491\atop -0.48975}$ &
$\displaystyle{0.69490\atop -0.48972}$ &
$\displaystyle{0.69491\atop -0.48975}$ \\\\

${\displaystyle-\frac{\pi}{8}}$ & $\displaystyle{R^{\mathstrut}e\atop
I_{\mathstrut}m}$ &
$\displaystyle{12.3\atop -7.5}$ &
$\displaystyle{0.81233\atop -0.66767}$ &
$\displaystyle{0.81233\atop -0.66767}$ &
$\displaystyle{0.81231\atop -0.66764}$ &
$\displaystyle{0.81233\atop -0.66767}$ \\\\

${\displaystyle-\frac{\pi}{4}}$ & $\displaystyle{R^{\mathstrut}e\atop
I_{\mathstrut}m}$ &
$\displaystyle{{-2\,\ (11)}\atop {3\,\ (12)}}$ &
$\displaystyle{-87\atop 47}$ &
$\displaystyle{1.0360\atop -0.85549}$ &
$\displaystyle{1.0360\atop -0.85546}$ &
$\displaystyle{1.0360\atop -0.85549}$ \\\\

${\displaystyle-\frac{3\pi}{8}}$ & $\displaystyle{R^{\mathstrut}e\atop
I_{\mathstrut}m}$ &
$\displaystyle{{9\,\ (17)}\atop {9\,\ (19)}}$ &
$\displaystyle{{3\,\ (12)}\atop {-2\,\ (13)}}$ &
$\displaystyle{{1\,\ (4)}\atop {3\,\ (4)}}$ &
$\displaystyle{1.5184\atop -1.1802}$ &
$\displaystyle{1.5185\atop -1.1802}$ \\
\end{tabular}
\end{table}


\begin{table}
\caption[]{
 Values of $F_0^{(+)}$ and $F_0^{(-)}$ for different $x$.
The values of $k$ and $\theta$ are fixed at
 $2\ e^{i\frac{\pi}{4}} \,{\rm fm.}^{-1}$ and 0 respectively.
The numbers in brackets give the exponent power of ten.}
\vskip 0.2cm
\begin{tabular}{c@{\hspace{8mm}}c@{\hspace{8mm}}c@{\hspace{8mm}}c}
$x\,({\rm fm})$ & $F_0^{(+)}(k,x,0)$ & $F_0^{(-)}(k,x,0)$ &
${{\left|\displaystyle{\frac{h_0^{(+)}F_0^{(+)}}{h_0^{(-)}F_0^{(-)}}}
\right|}_{\displaystyle{\mathstrut}}}^{\displaystyle{\mathstrut}}$\\
\tableline
$0.5^{\mathstrut}_{\mathstrut}$ & $0.62+i\ 0.23$ & $0.82292 -i\ 0.06269$ &
0.20 \\

$1^{\mathstrut}_{\mathstrut}$ & $0.59+i\ 1.40$ & $0.70057 -i\ 0.15146$ &
0.13 \\

$2^{\mathstrut}_{\mathstrut}$ & $3.30-i\ 2.24$ & $0.62144 -i\ 0.20521$ &
0.02 \\

$3^{\mathstrut}_{\mathstrut}$ & $-12.4+i\ 11.6$ & $0.59993 -i\ 0.21596$ &
0.6\,\ (-2) \\

$5^{\mathstrut}_{\mathstrut}$ & $-353+i\ 23.8$ & $0.59261 -i\ 0.21939$ &
0.4\,\ (-3) \\

$10^{\mathstrut}_{\mathstrut}$ & $5\ (4) +i\ 7\ (5)$ & $0.59205-i\ 0.21965$ &
0.6\,\ (-6) \\

$15^{\mathstrut}_{\mathstrut}$ & $(9)-i\ (8)$ & $0.59205 -i\ 0.21965$ &
0.8\,\ (-9) \\

$20^{\mathstrut}_{\mathstrut}$ & $-2\ (11)-i\ 3\ (12)$ & $0.59205
-i\ 0.21965$ & 0.1\ (-11) \\
\end{tabular}
\end{table}

\newpage

\begin{figure}
\centering
\unitlength=0.70mm
\special{em:linewidth .5pt}
\linethickness{.5pt}
\begin{picture}(129.67,129.67)
       \put({32.33},{129.67}){\special{em:moveto}}
       \put({32.33},{25.00}){\special{em:lineto}}
       \put({7.00},{80.67}){\special{em:moveto}}
       \put({129.67},{80.67}){\special{em:lineto}}
\put(32.33,93.67){\circle{3.33}}
       \put({32.33},{80.67}){\special{em:moveto}}
       \put({97.33},{64.00}){\special{em:lineto}}
\put(32.33,105.67){\circle{3.33}}
\put(117.33,46.33){\circle*{3.33}}
       \put({7.33},{116.67}){\special{em:moveto}}
       \put({70.33},{26.67}){\special{em:lineto}}
\put(32.33,93.67){\circle*{3.33}}
\put(32.33,105.33){\circle*{3.33}}
\put(32.33,118.67){\circle*{3.33}}
\put(98.00,63.67){\circle*{3.33}}
       \put({71.72},{80.67}){\special{em:moveto}}
       \put({72.27},{79.62}){\special{em:lineto}}
       \put({72.27},{79.62}){\special{em:moveto}}
       \put({72.64},{78.58}){\special{em:lineto}}
       \put({72.64},{78.58}){\special{em:moveto}}
       \put({72.82},{77.47}){\special{em:lineto}}
       \put({72.82},{77.47}){\special{em:moveto}}
       \put({72.89},{76.21}){\special{em:lineto}}
       \put({72.89},{76.21}){\special{em:moveto}}
       \put({72.79},{75.29}){\special{em:lineto}}
       \put({72.79},{75.38}){\special{em:moveto}}
       \put({72.64},{74.33}){\special{em:lineto}}
       \put({72.64},{74.33}){\special{em:moveto}}
       \put({72.39},{73.44}){\special{em:lineto}}
       \put({72.39},{73.44}){\special{em:moveto}}
       \put({71.99},{72.30}){\special{em:lineto}}
       \put({71.99},{72.30}){\special{em:moveto}}
       \put({71.56},{71.51}){\special{em:lineto}}
       \put({71.56},{71.51}){\special{em:moveto}}
       \put({71.19},{70.98}){\special{em:lineto}}
       \put({71.19},{70.98}){\special{em:moveto}}
       \put({71.01},{70.77}){\special{em:lineto}}
       \put({44.32},{80.67}){\special{em:moveto}}
       \put({44.32},{79.49}){\special{em:lineto}}
       \put({44.32},{79.49}){\special{em:moveto}}
       \put({44.23},{78.13}){\special{em:lineto}}
       \put({44.23},{78.13}){\special{em:moveto}}
       \put({44.10},{77.40}){\special{em:lineto}}
       \put({44.10},{77.40}){\special{em:moveto}}
       \put({43.92},{76.49}){\special{em:lineto}}
       \put({43.92},{76.49}){\special{em:moveto}}
       \put({43.64},{75.58}){\special{em:lineto}}
       \put({43.64},{75.58}){\special{em:moveto}}
       \put({43.37},{74.67}){\special{em:lineto}}
       \put({43.37},{74.67}){\special{em:moveto}}
       \put({42.92},{73.67}){\special{em:lineto}}
       \put({42.92},{73.67}){\special{em:moveto}}
       \put({42.33},{72.77}){\special{em:lineto}}
       \put({42.33},{72.77}){\special{em:moveto}}
       \put({41.64},{71.81}){\special{em:lineto}}
       \put({41.64},{71.81}){\special{em:moveto}}
       \put({40.92},{71.04}){\special{em:lineto}}
       \put({40.92},{71.04}){\special{em:moveto}}
       \put({40.24},{70.45}){\special{em:lineto}}
       \put({40.24},{70.45}){\special{em:moveto}}
       \put({39.78},{70.27}){\special{em:lineto}}
       \put({40.51},{69.31}){\special{em:moveto}}
       \put({41.51},{69.99}){\special{em:lineto}}
       \put({41.51},{69.99}){\special{em:moveto}}
       \put({42.60},{71.13}){\special{em:lineto}}
       \put({42.60},{71.13}){\special{em:moveto}}
       \put({43.60},{72.67}){\special{em:lineto}}
       \put({43.60},{72.67}){\special{em:moveto}}
       \put({44.42},{74.40}){\special{em:lineto}}
       \put({44.42},{74.40}){\special{em:moveto}}
       \put({44.87},{75.76}){\special{em:lineto}}
       \put({44.87},{75.76}){\special{em:moveto}}
       \put({45.01},{76.67}){\special{em:lineto}}
       \put({45.01},{76.67}){\special{em:moveto}}
       \put({45.19},{77.72}){\special{em:lineto}}
       \put({45.19},{77.72}){\special{em:moveto}}
       \put({45.28},{78.72}){\special{em:lineto}}
       \put({45.28},{78.72}){\special{em:moveto}}
       \put({45.32},{79.62}){\special{em:lineto}}
       \put({45.32},{79.62}){\special{em:moveto}}
       \put({45.32},{80.67}){\special{em:lineto}}
\put(115.67,85.00){\makebox(0,0)[lc]{$R\!e\,k$}}
\put(69.33,31.67){\makebox(0,0)[lc]{dividing line}}
\put(76.67,75.00){\makebox(0,0)[lc]{$\varphi$}}
\put(45.67,69.00){\makebox(0,0)[lc]{$\theta$}}
\put(35.33,126.33){\makebox(0,0)[lb]{$I\!m\,k$}}
\put(36.00,118.00){\makebox(0,0)[lc]{$k_{01}$}}
\put(36.00,105.00){\makebox(0,0)[lc]{$k_{02}$}}
\put(36.00,93.00){\makebox(0,0)[lc]{$k_{03}$}}
\put(102.33,63.67){\makebox(0,0)[lc]{$k_{04}$}}
\put(122.00,46.00){\makebox(0,0)[lc]{$k_{05}$}}
\end{picture}
\caption{ Schematical picture of a typical distribution of the spectral
points $k_{0i}$ on the complex $k$-plane. The dividing line defined
by the rotation angle $\theta$ is also shown.}
\label{pic1}
\end{figure}
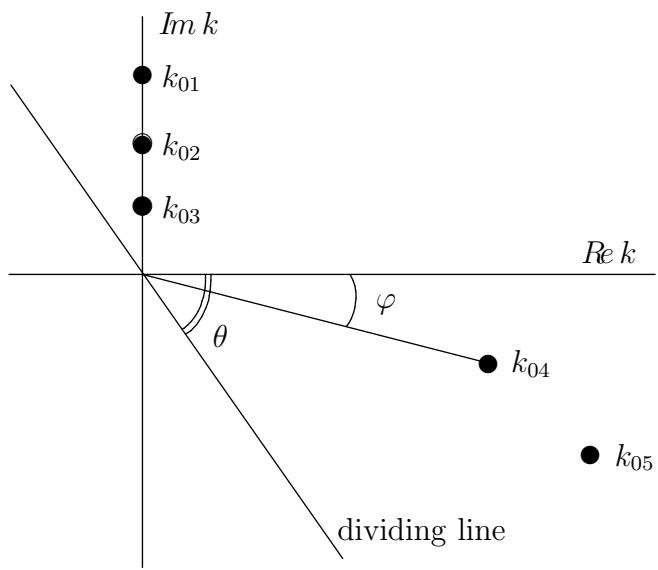

\newpage

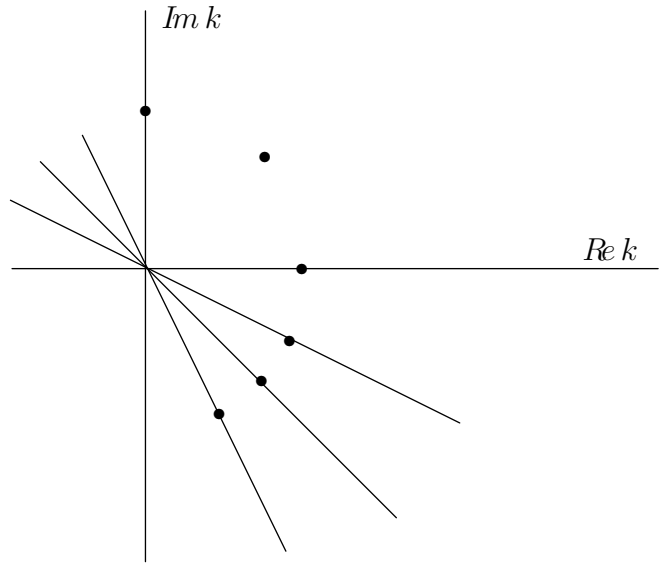
\begin{figure}
\centering
\unitlength=0.70mm
\special{em:linewidth .5pt}
\linethickness{.5pt}
\begin{picture}(129.67,129.67)
       \put({32.33},{129.67}){\special{em:moveto}}
       \put({32.33},{25.00}){\special{em:lineto}}
       \put({7.00},{80.67}){\special{em:moveto}}
       \put({129.67},{80.67}){\special{em:lineto}}
\put(115.34,84.33){\makebox(0,0)[lc]{$R\!e\,k$}}
\put(35.33,126.33){\makebox(0,0)[lb]{$I\!m\,k$}}
       \put({12.33},{101.00}){\special{em:moveto}}
       \put({80.00},{33.33}){\special{em:lineto}}
       \put({92.00},{51.33}){\special{em:moveto}}
       \put({6.67},{93.67}){\special{em:lineto}}
       \put({59.00},{27.00}){\special{em:moveto}}
       \put({20.33},{106.00}){\special{em:lineto}}
\put(59.67,67.00){\circle*{1.89}}
\put(54.33,59.33){\circle*{1.89}}
\put(46.33,53.00){\circle*{1.89}}
\put(62.00,80.67){\circle*{1.89}}
\put(32.33,110.67){\circle*{1.89}}
\put(55.00,102.00){\circle*{1.89}}
\end{picture}
\caption{ Several dividing lines
corresponding to the Jost function calculations given in Table I.
The points at which this function was calculated  are shown by the dots.}
\label{pic2}
\end{figure}

\end{document}